\documentclass[12pt]{iopart}

\usepackage{graphicx}
\usepackage{hyperref}

\begin{document}

\title{An ytterbium quantum gas microscope with narrow-line laser cooling}

\author{Ryuta Yamamoto, Jun Kobayashi, Takuma Kuno, Kohei Kato, Yoshiro Takahashi}
\address{Department of Physics, Graduate School of Science, Kyoto University, Japan 606-8502}
\ead{r\_yamamoto@scphys.kyoto-u.ac.jp}

\begin{abstract}
We demonstrate site-resolved imaging of individual bosonic $^{174}\mathrm{Yb}$ atoms
in a Hubbard-regime two-dimensional optical lattice with a short lattice constant of 266 nm.
To suppress the heating by probe light with the $^1S_0$-$^1P_1$ transition of the wavelength
$\lambda$ = 399 nm for high-resolution imaging
and preserve atoms at the same lattice sites during the fluorescence imaging,
we simultaneously cool atoms by additionally applying narrow-line optical molasses
with the $^1S_0$-$^3P_1$ transition of the wavelength $\lambda$ = 556 nm.
We achieve a low temperature of $T = 7.4(1.3)\ \mu\mathrm{K}$,
corresponding to a mean oscillation quantum number along the horizontal axes of 0.22(4)
during imaging process.
We detect on average 200 fluorescence photons from a single atom within 400 ms exposure time,
and estimate the detection fidelity of 87(2)\%.
The realization of a quantum gas microscope with enough fidelity for Yb atoms
in a Hubbard-regime optical lattice
opens up the possibilities for studying various kinds of quantum many-body systems
such as Bose and Fermi gases, and their mixtures,
and also long-range-interacting systems such as  Rydberg states.
\end{abstract}

\pacs{67.85.Hj, 07.60.Pb, 37.10.Jk, 37.10.De}
\vspace{2pc}
\noindent{\it Keywords}: cold atoms, optical lattice, laser cooling, quantum gas microscope

\maketitle

\section{Introduction}
Ultracold quantum gases in optical lattices have proven extremely useful for the study of quantum phases
and the dynamical evolutions of strongly correlated many-body system described by a Hubbard model \cite{Zoller:HubbardToolbox2005}.
Well-known examples include a quantum phase transition from a superfluid (SF) to a Mott insulator (MI) for bosonic species
\cite{Bloch:SFMI2002,Esslinger:SFMI(1D)2004,Porto:SFMI(2D)2007},
and a crossover from a metal to a MI for fermionic species
\cite{Rosch:MetalMI2008,Esslinge:MetalMI2008}.
To fully exploit the potential of ultracold atoms in an optical lattice as a quantum simulator,
it is great advantage to have access to the in-trap atom distribution with single-atom resolution.
In 2009, a quantum gas microscope (QGM) has been realized with bosonic $^{87}\mathrm{Rb}$
for the first time
\cite{Greiner:Rb-QGM2009}.
Site-resolved imaging has been used to study the SF to MI transition
\cite{Bloch:Rb-QGM:Mott2010,Greiner:Rb-QGM:Mott2010},
and strongly correlated dynamics in optical lattices
\cite{Greiner:Rb-QGM:QW2015,Bloch:Rb-QGM:SpinInpurity2013,Bloch:Rb-QGM:LCSpread2012}.
Site-resolved imaging systems have been realized for other alkaline atomic species
such as fermionic $^{40}\mathrm{K}$ \cite{Zweirlein:K-QGM2015,Kuhr:K-QGM2015} and
$^{6}\mathrm{Li}$ \cite{Greiner:Li-QGM2015} very recently.	

Extending the applicability of a QGM technique to atomic species beyond alkali-metal atoms is an important step for a further revolution.
In particular, a successful application of a QGM technique to two-electron atoms such as alkali-earth metal
and ytterbium (Yb) atoms is remarkable because it offers many unique possibilities for the quantum simulation and quantum information researches.
In fact, recent studies demonstrate	 that a system of two-electron atoms in an optical lattice is an ideal platform for the study of SU($\mathcal{N}$) physics
\cite{Cazalilla:SU(N)2009, STaie:SU(6)2012, AMRey:SU(N)2012, Fallani:1DSU(N)2014},
two-orbital SU($\mathcal{N}$) physics
\cite{Gorshkov:Two-orbitSU(N)2010, Folling:two-orbitalSU(N)2014, Ye:SpectroscopySU(N)2014, Fallani:SpinExchange2014},
and topological physics
\cite{Fallani:TopologicalEdge2015}.
In addition, a variety of stable isotopes, 5 bosonic and 2 fermionic isotopes in the case of Yb atoms for example,
we can study various kinds of many-body systems such as ultracold Bose and Fermi gases
and  Bose-Bose
\cite{SSugawa:168YbBEC, Schreck:SrDegenerateMixture2013}, Bose-Fermi
\cite{Schreck:SrDegenerateMixture2013, Schreck:SrBFM2010, SSugawa:YbBFM2011},
and Fermi-Fermi
\cite{STaie:SU(6)2012} mixtures in an optical lattice.
The existence of nuclear spin degrees of freedom in the ground state $^1S_0$ and
long-lived metastable states $^3P_0$ and $^3P_2$ offers unique possibilities for quantum memory and quantum computation
\cite{Derevianko:QC2004, Daley:QC2008, Gorshkov:QuantumRegister2009, KShibata:QC2009}.
Additionally we can tune interatomic interactions between the $^1S_0$ and $^3P_2$ states  by an anisotropy-induced magnetic Feshbach resonance
\cite{SKato:Yb3P2Feshbach}.
Furthermore, a high-resolution laser spectroscopy of atoms in an optical lattice using the ultranarrow $^1S_0$-$^3P_0$ and $^1S_0$-$^3P_2$ optical transitions
is also demonstrated both for bosons and fermions, revealing the novel behavior of the atomic interaction of the system
\cite{Folling:two-orbitalSU(N)2014, Ye:SpectroscopySU(N)2014, Fallani:SpinExchange2014, SKato:Yb3P2Feshbach, AYamaguchi:3P2Spectroscopy2010}.
There has been also considerable interest in high-lying Rydberg states of  two-electron atoms
\cite{Jones:SrRydberg2010, Killian:SrRydberg2013}
in an optical lattice
\cite{Jones:RydbergOL2011}
because of an additional degrees of freedom for probing and manipulation provided by  the   remaining valence electron of a singly excited Rydberg state.  
The successful application of a QGM technique to these systems will definitely enhance our understanding of the physics behind.  

The important progress towards this direction has been reported quite recently in ref. \cite{Kozuma:Yb-QGM2015},
in which a site-resolved imaging system has been realized without cooling process for $^{174}\mathrm{Yb}$ atoms in a two-dimensional (2D) optical lattice with a lattice constant of 544 nm.
The achieved resolution of $\sim 310$ nm (full width at half maximum, FWHM) is impressively small.
A further study is still necessary, however, to successfully perform the above mentioned interesting researches for Yb atoms using a QGM.
First, a crucial aspect of QGM is the high-fidelity of the imaging process characterized by loss and hopping rates during the fluorescence imaging,
which should be evaluated by comparing two successive images taken for the same atoms.
Second, the Hubbard-regime optical lattice needs a shorter lattice constant, especially for heavier atoms of Yb.   
These conditions should be  simultaneously satisfied with
the single-site resolved imaging and single-atom sensitivity. 

In this work, we achieve site-resolved imaging of individual $^{174}\mathrm{Yb}$ atoms in a 2D optical lattice
with a short lattice constant of 266 nm which ensures the Hubbard-regime \cite{TFukuhara:174YbSF-MI2009}.
To keep atoms at the same lattice sites during the fluorescence imaging, 
we simultaneously cool atoms by additionally applying narrow-line optical molasses with the $^1S_0$-$^3P_1$ transition
($\lambda = 556$ nm, the Doppler limit temperature $T_D = 4.4\ \mu\mathrm{K}$, natural linewidth $\Gamma / 2\pi = 182\ \mathrm{kHz}$),
resulting in a low temperature of $T = 7.4(1.3)\ \mu\mathrm{K}$,
corresponding to a mean oscillation quantum number along the horizontal axes 0.22(4)
during imaging process
(see \fref{fig:ImagingSystem}(a) for relevant energy levels).
In particular, the careful tuning of the relative angle between an applied magnetic field and a polarization of lattice laser beams realizes the cancellation of the inhomogeneity of the light-shifts,
which enhances a cooling efficiency of narrow-line laser cooling, both for sideband cooling along the horizontal direction and Doppler cooling along the vertical direction.
The realization of such high efficient cooling makes possible
to suppress the heating due to the probe light using the $^1S_0$-$^1P_1$ transition
($\lambda = 399$ nm, the Doppler limit temperature $T_D = 690\ \mu\mathrm{K}$, natural linewidth $\Gamma / 2\pi = 29\ \mathrm{MHz}$) for high-resolution imaging
(see \fref{fig:ImagingSystem}(a)).
We achieve a  lifetime $\tau > 7$ s of atoms during fluorescence imaging much longer than a typical imaging time of 400 ms,
enabling to take multiple images for the same atomic sample and to successfully estimate the imaging fidelity to be 87(2)\%. 
The realization of a QGM
with enough fidelity for Yb atoms in a Hubbard-regime optical lattice opens up the possibilities for studying various kinds of quantum many-body systems
such as Bose and Fermi gases, and their mixtures in an optical lattice, and also long-range-interacting systems such as Rydberg states.

The paper is organized as follow.
In \sref{sec:Setup}, we describe an experimental setup
and a method for preparation of an atom sample in a 2D lattice system.
\Sref{sec:Narrow-line laser cooling in an optical lattice} presents cooling mechanism for realization of a site-resolved imaging
and \sref{sec:Site-resolved imaging} shows the results of the analysis of site-resolved fluorescence images including loss and hopping rates.
Finally we conclude our work with brief prospects in \sref{sec:Conclusion}.

\section{Experimental setup and atom preparation}
\label{sec:Setup}
\begin{figure}[tb]
\begin{minipage}[b]{0.45\linewidth}
\centering
\includegraphics{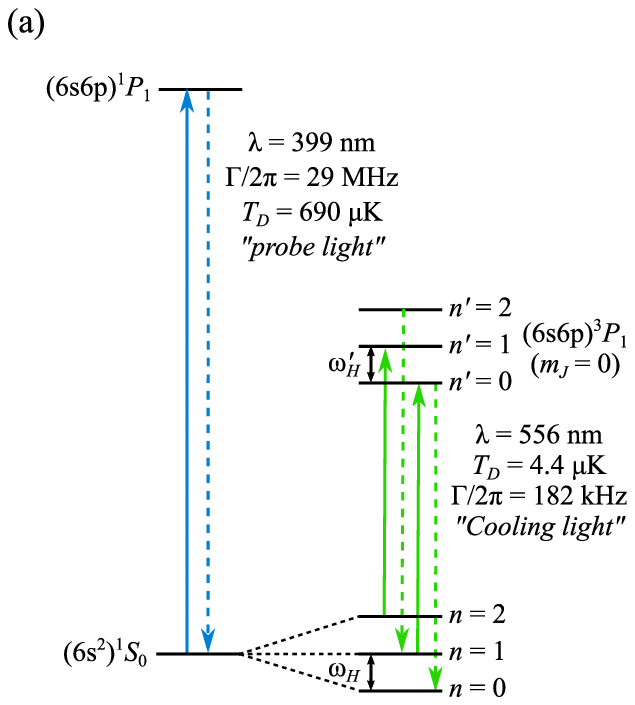}
\label{figure1a}
\end{minipage}
\begin{minipage}[t]{0.55\linewidth}
\centering
\includegraphics{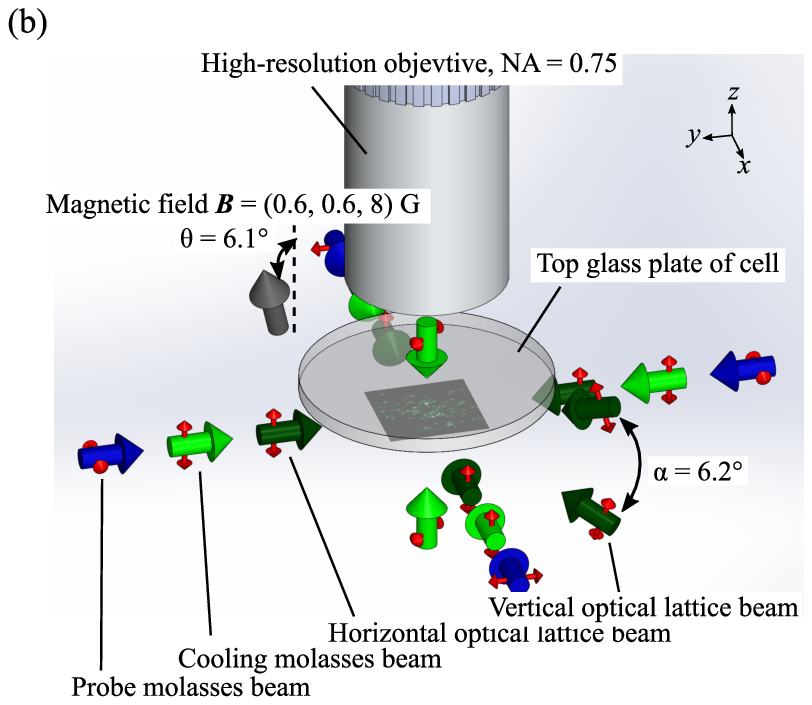}
\label{figure1b}
\end{minipage}
\caption{
(a) Low-lying energy levels of $^{174}\mathrm{Yb}$ atoms relevant for imaging and cooling scheme.
The $^1S_0$-$^1P_1$ transition
is used for high-resolution imaging and the $^1S_0$-$^3P_1$ transition
for high efficient cooling.
$n$ and $n^\prime$ show the vibrational level of the $^1S_0$ and $^3P_1(m_J = 0)$ states, respectively.
$\omega_H$ and $\omega^\prime_H$ show the trap frequency
along the horizontal lattice of the $^1S_0$ and $^3P_1(m_J = 0)$ states, respectively.
(b) Experimental setup for high-resolution imaging in a deep optical lattice.
Dark and light green and blue arrows show the direction of lattice, 556 nm cooling molasses, and 399 nm probe molasses beams, respectively.
Red arrows show the polarization of each laser beam.
The high-resolution objective with NA = 0.75
is just above the glass cell, made of plates with 3 mm thickness.
The wavelength of all lattice beams is 532 nm
and the beam waists of lattice beams along the x, y, and z directions
are $(w_x, w_y, w_z) \cong (23, 23, 15)\ \mu\mathrm{m}$, respectively.
}
\label{fig:ImagingSystem}
\end{figure}

Our experiment starts from loading $^{174}\mathrm{Yb}$ atoms into a magneto-optical trap (MOT) in a metal chamber
and transferring the atoms into a glass cell (Schott AG BOROFLOAT) using an optical tweezer (OT).
The detail of the MOT and OT setup is described in ref. \cite{SKato:OMRI2012}.
The position of atoms in the glass cell is
about 5.5 mm below the surface of the glass cell,
and just 6.23 mm under a high-resolution objective with numerical aperture of NA = 0.75
(Mitutoyo G Plan Apo HR50x (custom)), which is schematically shown in \fref{fig:ImagingSystem}(b). 
After creating a Bose Einstein Condensate (BEC) of $5 \times 10^4$ after 10 s evaporative cooling in a crossed optical trap formed by the OT beam and another 532 nm beam,
we load the BEC into a vertical lattice generated by the interference of two laser beams
with the wavelength of $\lambda = 532\ \mathrm{nm}$ propagating
at a relative angle of $\alpha = 6.2^\circ$.
The vertical lattice has a spacing of $\lambda/2\sin(\alpha/2)=4.9\ \mu\mathrm{m}$
and the trap frequency along the vertical axis (z-axis) of $\omega_z = 2\pi \times 2\ \mathrm{kHz}$ at this loading stage,
as explained in detail in our previous work \cite{KShibata:Yb-QGM2014}.

\begin{figure}[tb]
\centering
\includegraphics{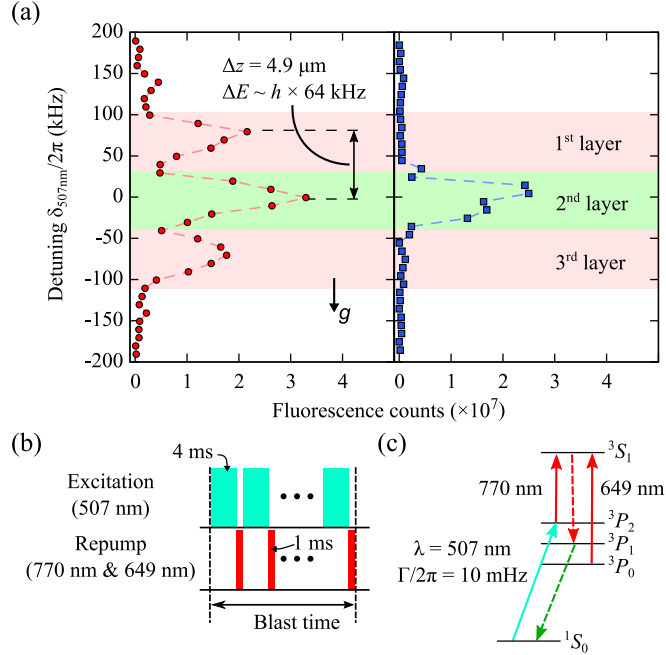}
\caption{
(a) Spectroscopy of atoms in the vertical lattice using the $^1S_0$-$^3P_2(m_J=-1)$ transition
at a bias magnetic field $B_z = 1.4$ G
and a magnetic field gradient $\Delta B = 7\ \mathrm{mG/\mu m}$.
The left and right panels show the spectrum without and with blasting away atoms
in the additional layers, respectively.
A layer separation is $\Delta z = 4.9\ \mu\mathrm{m}$
corresponding to the Zeeman shift $\Delta E_{\mathrm{Zeeman}} = h \times 70\ \mathrm{kHz}$
which is in agreement with the measured energy separation $\Delta E$ of $h \times 64\ \mathrm{kHz}$.
The arrow with $g$ shows the direction of gravity.
(b) Sequence of preparing only atoms in a single-layer ($2^\mathrm{nd}$ layer).
We blast away the atoms trapped in the $\mathrm{1^{st}}$ and $\mathrm{3^{rd}}$ layers
by alternately exciting the atoms into $^3P_2$ state followed by the rapid inelastic collision decay
and repumping back them into $^1S_0$ state.
The blast time is typically 250 ms.
(c) Low-lying energy levels of $^{174}\mathrm{Yb}$ atoms relevant for blast.
}
\label{fig:LayerBlast}
\end{figure}

The atoms just after loaded into the vertical lattice spread over several, typically three, layers, as shown in the left panel of \fref{fig:LayerBlast}(a).
In this situation, although we can focus on the atoms in one selected layer with an objective depth of less than $1\ \mu \mathrm{m}$,
we have always contributions from the atoms in other layers which considerably blur the image.
To observe clear image for the atoms in only one layer, we blast away the atoms in unnecessary layers.
This is done by alternately exciting the atoms into the $^3P_2(m_J = -1)$ state
under a bias magnetic field $B = 1.4\ \mathrm{G}$ and
a magnetic field gradient $\Delta B = 7\ \mathrm{mG/\mu m}$,
corresponding to the Zeeman shift $\Delta E_{\mathrm{Zeeman}} = h\times 70\ \mathrm{kHz/layer}$,
followed by the rapid inelastic collisional decay and repumping back the atoms into the ground state (see \fref{fig:LayerBlast}(b)).
Here the ultranarrow optical transition of $^1S_0$-$^3P_2(m_J = -1)$ with the resonant wavelength of $507\ \mathrm{nm}$
and natural linewidth $\Gamma/2\pi = 10 \mathrm{mHz}$ is used for the excitation and $^3P_2(m_J = -1)$-$^3S_1$
and $^3P_0$-$^3S_1$ for repumping (see \fref{fig:LayerBlast}(c)). 
As a result, we successfully prepare the atoms in only a single layer, as shown in the right panel of \fref{fig:LayerBlast}(a).

Finally, we load the 2D atom cloud in a single layer into horizontal 2D optical lattices (x- and y-axes)
by simultaneously ramping up the potential of the vertical and horizontal optical lattices,
where the wavelength of horizontal lattices is $\lambda = 532\ \mathrm{nm}$ and the lattice spacing is
$266\ \mathrm{nm}$.
The vibrational frequencies along the three axes at the fluorescence imaging stage are 
$(\omega_x, \omega_y, \omega_z) = 2\pi \times (300, 300, 15.7)\ \mathrm{kHz}$, 
corresponding to the lattice depths of
$(U_x, U_y, U_z) = k_B \times (300, 300, 250)\ \mu \mathrm{K}$, respectively.

\section{Narrow-line laser cooling in an optical lattice}
\label{sec:Narrow-line laser cooling in an optical lattice}
An important prerequisite for realizing a QGM is to preserve the atoms at their sites during fluorescence imaging.
The $^1S_0$-$^1P_1$ transition provides high-resolution imaging with the diffraction limited resolution of a FWHM of 266 nm in our system.
However, the high Doppler cooling limit temperature $T_D = \mathrm{690}\ \mu \mathrm{K}$ makes quite difficult to preserve atoms at their sites.
In addition, the lack of hyperfine structure in the ground state $^1S_0$ of bosonic Yb atom makes impossible to apply sub-Doppler cooling techniques
such as polarization-gradient cooling and Raman sideband cooling.
To resolve this difficulty, we simultaneously cool the atoms with Doppler and sideband cooling using the $^1S_0$-$^3P_1$ narrow-line transition.

To efficiently cool all the atoms in an optical lattice with the narrow-line transition $^1S_0$-$^3P_1$,
we need to suppress the inhomogeneity of light shift
between the $^1S_0$ and $^3P_1$ states, otherwise the detuning for cooling is not optimized simultaneously for all atoms.
Here the light shift in the ground state $^1S_0$ is given by
$\Delta E_g^L = -(1/4) \alpha_g I$,
where $I$ is the laser intensity of the wavelength $\lambda = \mathrm{532\ nm}$
and $\alpha_g = 37.9\ \mathrm{Hz/(W/cm^2)}$ is
the calculated scalar polarizability in the $^1S_0$ state.
The light shift in the magnetic sublevel $m_J$ of the $^3P_1$ state is 
given as \cite{Rauschenbeutel:LightShift2013}
\begin{equation}
\label{eq:LightShift}
\Delta E_e^L = -\frac{1}{4}\alpha_e(m_J, \theta)\ I,
\end{equation}
where
\begin{equation}
\label{eq:Polarizabiliy}
\alpha_e(m_J, \theta) = \alpha_e^S - \alpha_e^T\frac{1 - 3 \cos^2\theta}{4}(m_J^2-2).
\end{equation}
Here $\theta$ is the angle between the quantization axis and the polarization of laser beams,
$\alpha_e^S$ and $\alpha_e^T$ are the scalar and tensor polarizabilities in the $^3P_1$ state, respectively.
Importantly,
equation \eref{eq:Polarizabiliy} provides the possibility of tuning the polarizability
$\alpha_e(m_J, \theta)$ to coincide with $\alpha_g$
by choosing an appropriate angle $\theta$ for $m_J$,
thus canceling the light-shift effects of the $^1S_0$-$^3P_1$ transition.
For this possibility we perform a laser spectroscopy with the $^1S_0$-$^3P_1$ transition for various laser intensities and angles $\theta$,
and accurately determine $\alpha_e^S$ and $\alpha_e^T$ as $22.4(2)$ and $-7.6(1)\ \mathrm{Hz/(W/cm^2)}$, respectively.
With these values, our current setup of the polarizations of all the lattice beams parallel to the vertical axis provides $\alpha_e(m_J=0, \theta=0)/\alpha_g = 0.99$.
In our experiment, however, we slightly tilt a magnetic field from the vertical direction
by an angle $6.1^\circ$ which gives $\alpha_e(m_J=0, \theta= 6.1^\circ)/\alpha_g = 0.98$.
This setup enables us to excite the atoms into the $^3P_1(m_J=0)$ state,
when the polarizations of the 556 nm cooling light along the horizontal axes are set to vertical,
and those along the vertical axis horizontal (see \fref{fig:ImagingSystem}(b)).
Note that the light shift of the $^1S_0$-$^1P_1$ transition
for probing is smaller than the natural linewidth of this transition of 29 MHz,
and so it is not a problem.
The total intensities of 399 nm and 556 nm beams correspond to the saturation parameters of $s_{399} \sim 1 \times 10^{-3}$ and $s_{556} \sim 1$, respectively.
With this dual molasses, Moir\'e patterns of about 6 $\mu \mathrm{m}$ pitch are observed as a result of the interference
between the cooling molasses beam of 556 nm and the optical lattice of 532 nm.  
To erase this unwanted Moir\'e pattern, we modulate the phase of the standing wave of the 556 nm optical molasses
by modulating retro-reflecting mirrors via the attached Piezo transducers,
as explained in detail in our previous work \cite{KShibata:Yb-QGM2014}.

\begin{figure}[tb]
\begin{minipage}[b]{0.5\linewidth}
\centering
\includegraphics{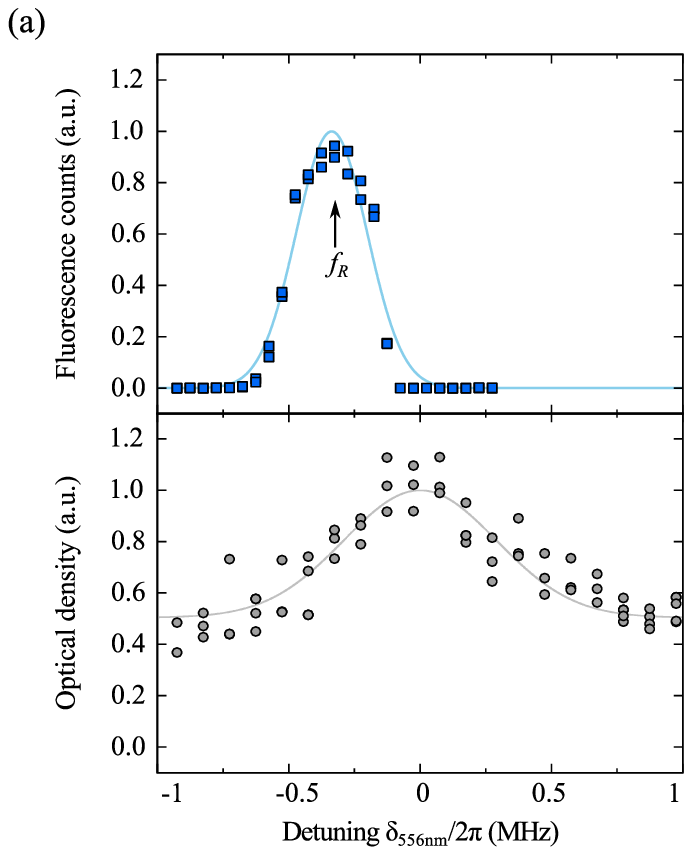}
\label{figure3a}
\end{minipage}
\begin{minipage}[b]{0.5\linewidth}
\centering
\includegraphics{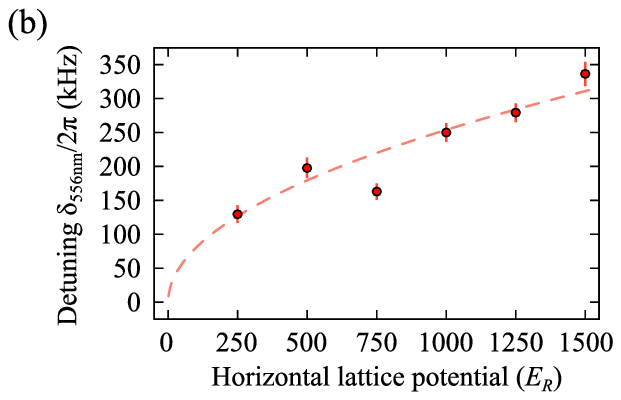}
\label{figure3b}
\includegraphics{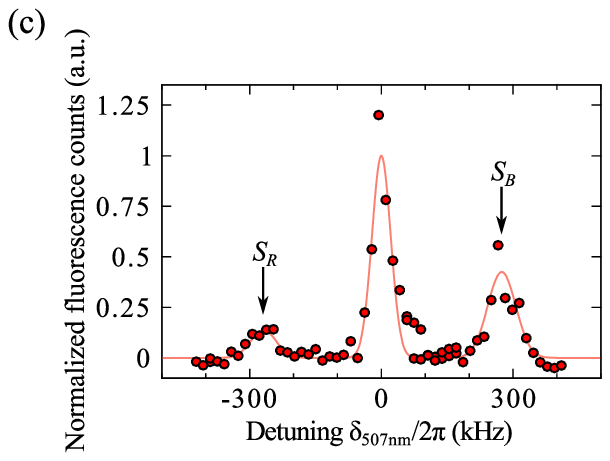}
\label{figure3c}
\end{minipage}
\caption{
(a) Spectra of atoms in a deep optical lattice with $U_x = U_y = 1500\ E_R$ as a function of the frequency of 556 nm detuning $\delta_{\mathrm{556}}/2\pi$.
The top panel (blue squares) shows the fluorescence counts of 399 nm probe molasses with cooling of 556 nm molasses beam along the horizontal axis.
The bottom panel (grey circles) shows the optical density measured by absorption imaging
with a 556 nm beam irradiated along the horizontal axis.
(b) Dependence of the optimal detuning of 556 nm cooling molasses on horizontal optical lattice depth.
Red circles show the experimental data.
Dotted red line shows the calculated trap frequency in an horizontal optical lattice.
(c) Laser spectroscopy of atoms using the $^1S_0$-$^3P_2\ (m_J=0)$ transition after cooling
by sideband (horizontal axis) and Doppler (vertical axis) cooling of 556 nm.
The ratio of the red to the blue sideband peaks $S_R/S_B$ is $0.32(6)$,
and the mean oscillation quantum number along horizontal axis
$\left\langle n \right\rangle = 0.22(4)$,
corresponding to the temperature along horizontal axis $T_H = 7.4(1.3)\ \mu\mathrm{K}$.
}
\label{fig:Sideband}
\end{figure}
The fine tuning of the relative angle between a magnetic field
and lattice laser polarizations indeed gives us a reasonably narrow resonance
of the $^1S_0$-$^3P_1(m_J=0)$ transition for atoms in the optical lattice
during fluorescence imaging.
\Fref{fig:Sideband}(a) shows the spectra of atoms
in our deepest horizontal optical lattices of $U_x = U_y = 1500\ E_R$,
where $E_R = h^2 / 2m\lambda_L^2 = k_\mathrm{B} \times 200\ \mathrm{nK}$
is recoil energy of lattice beam.
The top panel shows the fluorescence counts of 399 nm probe molasses light
as a function of a frequency of 556 nm cooling molasses beams along the horizontal axes,
in which we simultaneously apply the probe light and weak cooling molasses lights,
$s_{399} \sim 1\times10^{-3}$ and $s_{556} \sim 0.6$,
and we can observe many fluorescence counts during a 400 ms exposure time 
only when the cooling is efficient at favorable detunings.
We obtain the optimal frequency $f_R = -337(18)\ \mathrm{kHz}$
with the width of $318(12)$ kHz (FWHM).
The bottom panel shows the optical density measured by absorption imaging
with a 556 nm beam irradiated along the horizontal axis
as a function of a frequency of the 556 nm probe light.
In this measurement we do not apply 399 nm and 556 nm molasses beams.
Note that we set a zero frequency detuning as the resonance frequency of this spectrum.
We determine the optimal detuning of cooling beam along the horizontal axes
$\delta_{556}/2\pi= f_R = -337(18)\ \mathrm{kHz}$.
The same measurements are done at several horizontal lattice depths, as shown in \fref{fig:Sideband}(b).
In our lattice system, Lamb-Dicke parameters are
$\eta_\mu = \sqrt{\hbar k^2/(2m\omega_\mu)} = (0.11, 0.11, 0.48)$,
where $k$ is a wavevector of 556 nm light, and $\mu=x, y, z$.
Although the frequency separation between the cooling sideband $f_R$
and the main carrier $f=0$, corresponding to the trap frequencies $\omega_{x}$ and $\omega_{y}$,
is not large enough compared with the natural linewidth of 184 kHz for the 556 nm cooling transition, as shown in \fref{fig:Sideband}(a),
the responsible cooling mechanism along the x- and y-axes should be sideband cooling,
because the optimal detuning of cooling beam along the horizontal axis depends on the horizontal lattice trap depth
and is consistent with the trap frequency along the horizontal axis,
as shown in \fref{fig:Sideband}(b).

The temperature of the atoms during the fluorescence imaging is accurately measured by laser spectroscopy using the ultranarrow transition $^1S_0$-$^3P_2(m_J=0)$. 
\Fref{fig:Sideband}(c) shows the spectrum obtained,
which clearly shows three peaks corresponding to the red- and blue-sidebands and the main carrier.
From the ratio of the red to the blue sideband peaks $S_R/S_B = 0.32(6)$, we can evaluate a mean oscillation quantum number in a horizontal optical lattice 
$\langle n \rangle = \left(1 - S_R/S_B\right)^{-1/2} - 1 = 0.22(4)$
, corresponding to the atomic temperature of
$k_\mathrm{B}T_H = \hbar \omega / \ln{\left(1 + \langle n \rangle^{-1}\right)} = k_\mathrm{B} \times 7.4(1.3)\ \mu\mathrm{K}$
\cite{Wineland:LaserCooling1979, Wineland:Sideband1989}.
The value is in agreement with the theoretically predicted value of $6.4\ \mu \mathrm{K}$,
assuming narrow laser linewidth and no saturation.
The temperature along the vertical direction is also measured by a time-of-flight method
to be 12(1) $\mu \mathrm{K}$.
The optimal detunings are determined for various lattice depths,
and are independent of the lattice depth,
which suggests that the dominant cooling mechanism is Doppler cooling.
This is reasonable if we consider the small trap frequency of 15.7 kHz
along the vertical direction compared with the linewidth of 184 kHz.

\begin{figure}[tb]
\begin{minipage}[b]{0.5\linewidth}
\centering
\includegraphics{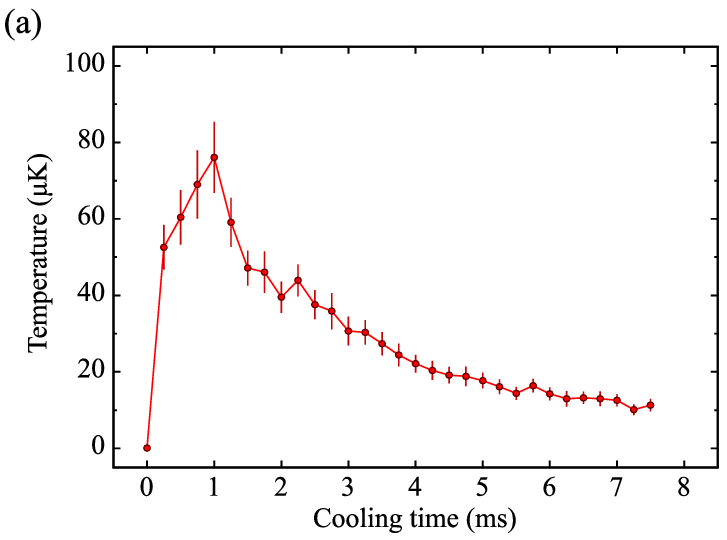}
\label{figure4a}
\end{minipage}
\begin{minipage}[b]{0.5\linewidth}
\centering
\includegraphics{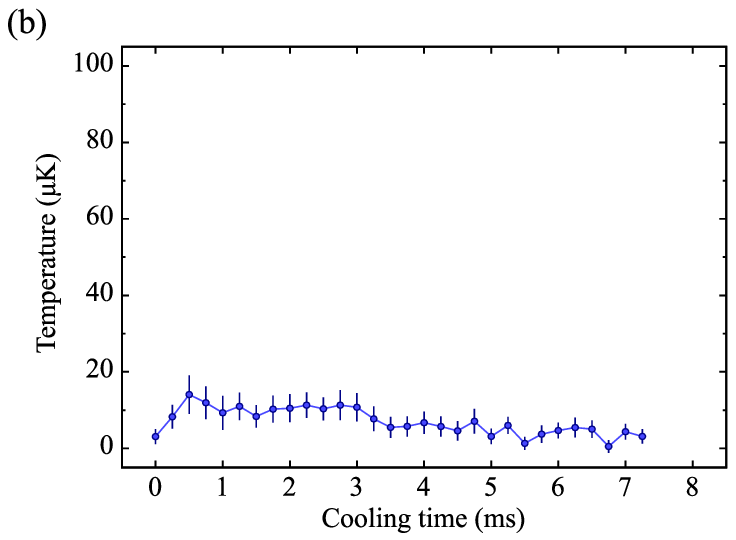}
\label{figure4b}
\end{minipage}
\caption{
Temporal evolution of the temperature (a) without and (b) with applying a PA pulse.
Temperatures in both graphs are measured by a time-of-flight method.
}
\label{fig:PADifferenceOfTemperature}
\end{figure}
We study the temporal evolution of the temperature during fluorescence imaging,
especially at the early stage of the imaging,
to investigate what happens during the imaging process.
In this measurement, the temperature is measured by a time-of-flight method with absorption imaging.
\Fref{fig:PADifferenceOfTemperature}(a) shows the results of the measurements.
The temperatures rapidly increase within several milliseconds
followed by the rather slow decease towards the steady-state value
obtained by the ultranarrow line laser spectroscopy of \fref{fig:Sideband}(c).
This behavior is explained as an effect of a light-assisted collision
due to the near-resonant probe and cooling light that
multiply-occupied atoms should be heated by a light-assisted collision
and subsequent release of the kinetic energy.
This is confirmed by further measurements with applying a photoassociation (PA) pulse
for removal of multiply-occupied sites before imaging.
\Fref{fig:PADifferenceOfTemperature}(b) shows that
the temperature remains a several $\mu \mathrm{K}$,
almost the same as the steady-state value,
which is consistent with no initial heating process as expected.
Although all the following single-site resolved imaging data presented in this paper
are measured without the application of PA light,
this initial heating effect is negligible 
because the multiply-occupied sites are almost absent in sparse atomic samples used for our QGM measurement.

\section{Site-resolved imaging}
\label{sec:Site-resolved imaging}
\begin{figure}[tb]
\begin{minipage}[b]{0.5\linewidth}
\centering
\includegraphics{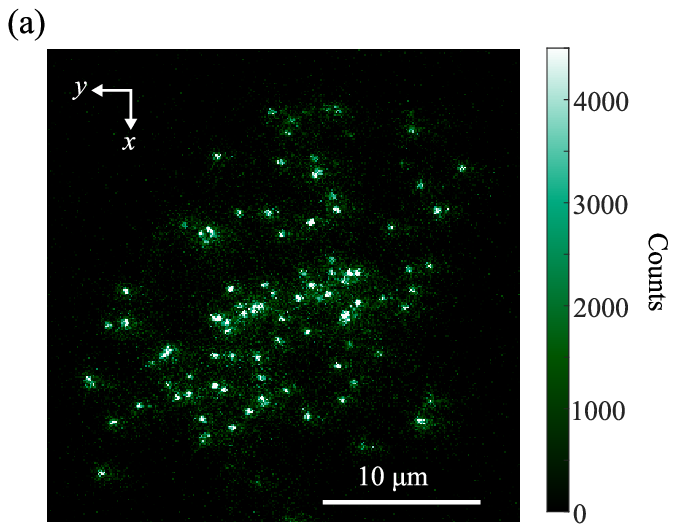}
\label{figure5a}
\end{minipage}
\begin{minipage}[b]{0.5\linewidth}
\centering
\includegraphics{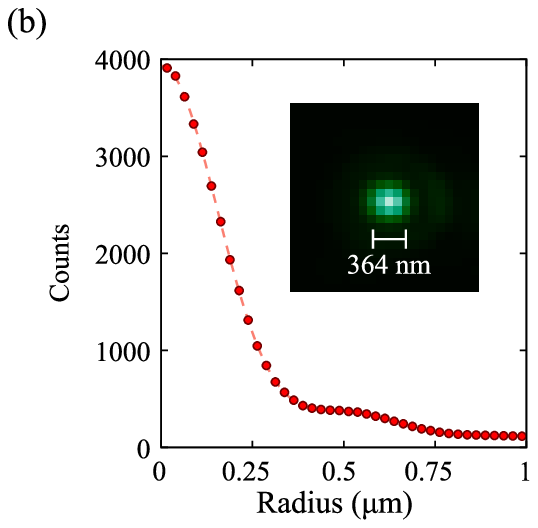}
\label{figure5b}
\end{minipage}\\
\begin{minipage}[c]{1\linewidth}
\centering
\includegraphics{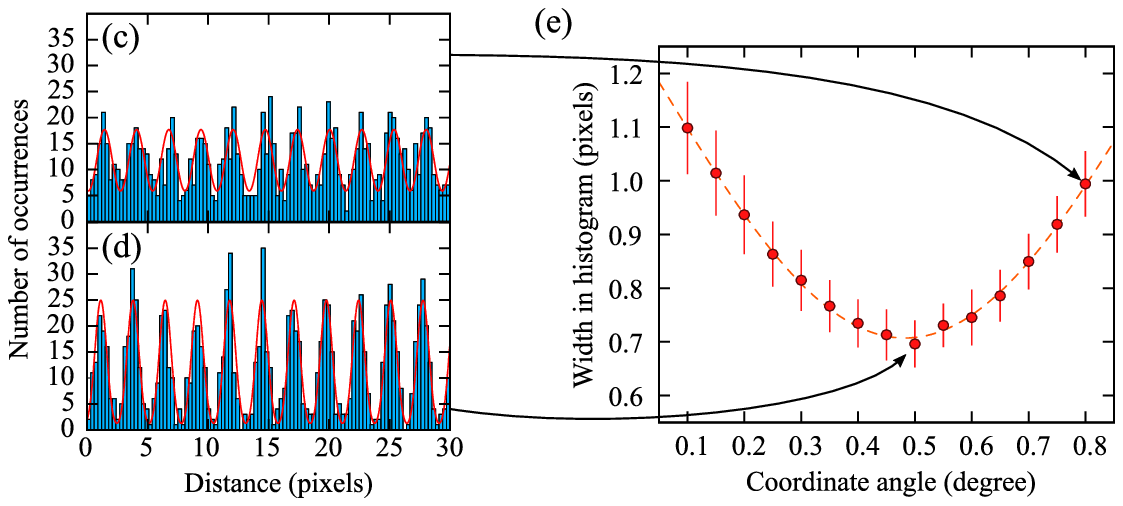}
\label{figure5c-5e}
\end{minipage}\\
\begin{minipage}[b]{1.0\linewidth}
\centering
\includegraphics{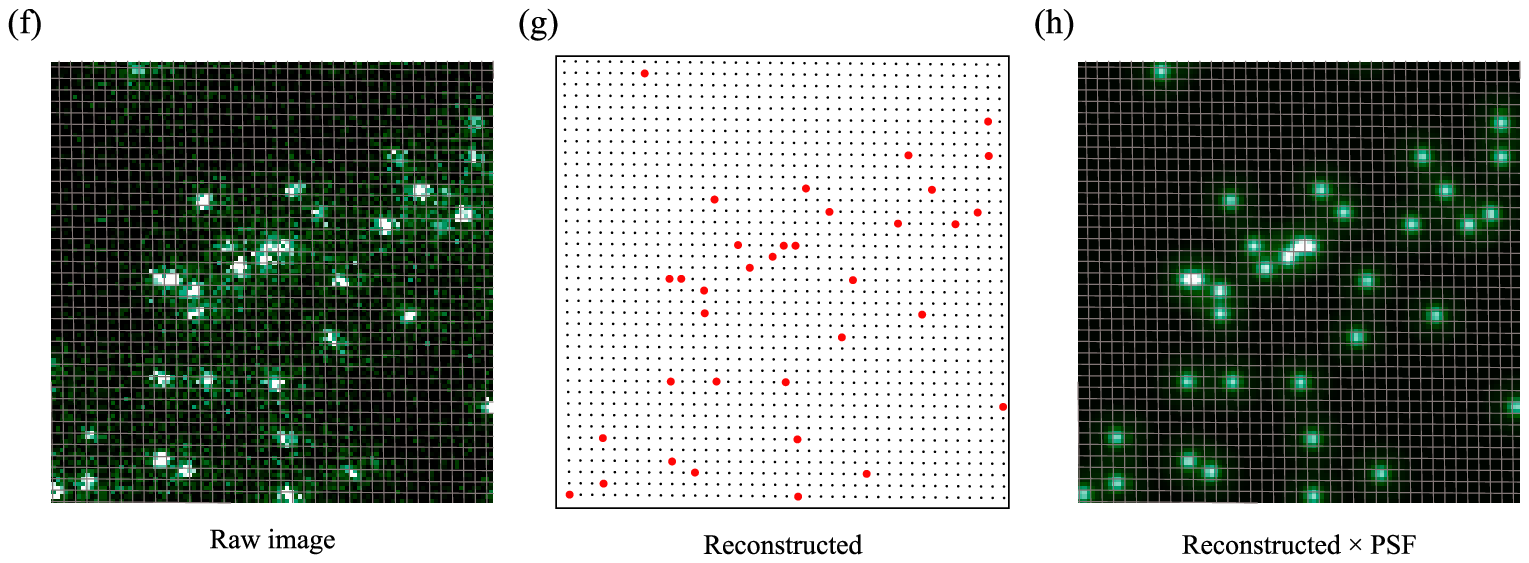}
\label{figure5f-5h}
\end{minipage}
\caption{
(a) Site-resolved imaging of $^{174}\mathrm{Yb}$ atoms on a sparsely-filled 266 nm-period optical lattice.
(b) The measured PSF averaged over $10^4$ individual single atoms
and azimuthal average of the PSF.
The line is a fit with a double Gaussian of equation \eref{eq:modelPSF} and yields
$A = 3850(10)$ counts,
$\sigma_1 = 154(1)\ \mathrm{nm}$,
$\sigma_2 = 153(10)\ \mathrm{nm}$,
$r_0 = 516(9)\ \mathrm{nm}$,
$B = 0.068(2)$, and
$C = 89(4)$ counts.
(c) and (d) The histogram of the mutual distances in the coordination rotated by an angle $\phi = 0.8^\circ$ and $0.5^\circ$, respectively. 
(e) Determination of lattice angle and spacing.
The red dotted line is a fit by
$\sigma(\phi) = \sigma_0 \left[1 + \beta \left\{1 - \cos\left(\gamma \left(\phi - \phi_0\right)\right)\right\}\right]$
and yields a minimum width at rotation angle of $\phi_0 = 0.482(3)^\circ$.
(f) Raw image of sparsely filled lattice
with grid lines showing lattice separation and orientation.
(g) Reconstructed atom distribution.
Red circles and black dots represent the atoms and the lattice sites, respectively.
(h) Numerically reconstructed atom distribution on lattice sites.
The image is convoluted with model PSF of equation \eref{eq:modelPSF} and reconstructed atom distribution of (g).
}
\label{fig:Image&PSF}
\end{figure}
We image the atomic fluorescence onto the EMCCD camera (Andor iXon${}^{\mathrm{EM}}$ Blue).
In \fref{fig:Image&PSF}(a) we show one illustrative example of the obtained images.
Note that, just before the fluorescence imaging, we intentionally select only about 2$\% $ of the atoms for easier evaluation of the performance of the QGM.
Such dilution of the atoms is done by performing a weak excitation with the $^1S_0$-$^3P_2(m_J=-1)$  transition, and then returning the atoms back into the ground state $^1S_0$. 
\Fref{fig:Image&PSF}(b) shows our measured point spread function (PSF),
obtained by averaging over $10^4$ fluorescence images of individual atoms.
We find that our PSF can be well approximated by a double Gaussian:
\begin{equation}
\label{eq:modelPSF}
PSF(r) = A \left[\exp\left(-\frac{r^2}{2\sigma_1^2}\right) + B \exp\left(-\frac{1}{2}\left(\frac{r-r_0}{\sigma_2}\right)^2\right)\right] + C
\end{equation}
with widths $\sigma_1,\ \sigma_2$, main and relative amplitudes $A,\ B$, a spatial offset $r_0$, and an overall count offset $C$.
The fit result shows our PSF is well described with $\sigma_1=154(1)$ nm and $\sigma_2=153(10)$ nm, and also has a FWHM of 364 nm,
and we detect on average 200 photons per atom within 400 ms fluorescence time.
Our system has a total fluorescence collection efficiency of 6.0\%,
given by the objective's solid angle of $\Omega/4\pi =$ 17\%,
51\% total transmission through the imaging optics,
and quantum efficiency of 70\% of our EMCCD camera.
The corresponding atomic fluorescence rate of $\sim 8300$ photons/s is large enough
to unambiguously identify the presence or absence of an atom for each lattice site.
Although the resolution of measured PSF is about 1.4 times larger than the lattice spacing,
we successfully determine the atomic distribution by deconvolution of images.

\begin{figure}[t]
\centering
\includegraphics{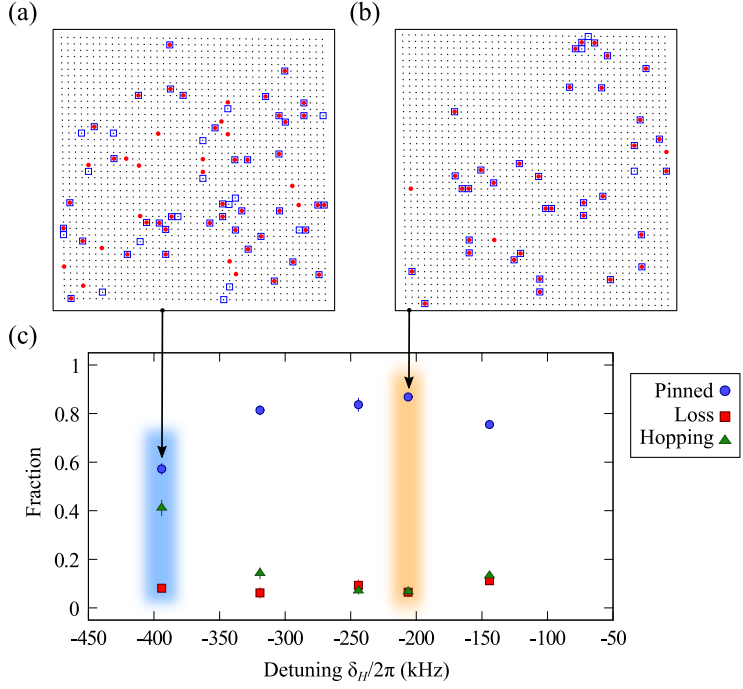}
\caption{
(a) and (b) Reconstructed atom distributions
at $\delta_{H}/2\pi = -394\ \mathrm{and} -206\ \mathrm{kHz}$, respectively.
Red circles and blue squares in the panels show the lattice sites occupation of the first and second image, respectively.
(c) The pinned, lost, and hopping fractions.
The fraction of pinned atoms (blue circles) shows the number of atoms preserved at the same lattice sites
in the two successive fluorescence images (400 ms exposure time, 300 ms delay between the two images).
The fraction of lost atoms (red squares) shows difference of the number of atoms between the two images.
The fraction of hopping atoms (green triangles) shows
the number of atoms appearing on a previously empty site in the second image.
All fractions are normalized to the number of atoms in the first image.
}
\label{fig:DistributionFraction}
\end{figure}
For reconstructing an atom distribution from our obtained images, 
we first determine a lattice angle and spacing
from the isolated, single-site resolved signals \cite{Bloch:Rb-QGM:Mott2010}.
Our lattice axes are oriented approximately along 
the vertical and horizontal axes with respect to the images.
The histogram of the mutual distances in the coordination rotated
by a small angle $\phi$ is shown in \ref{fig:Image&PSF}(c) and (d).
We fit a periodic array of Gaussians to the observed histogram.
\Fref{fig:Image&PSF}(e) shows the width of the Gaussians in the histogram
as a function of a coordinate rotation angle.
The red dotted line is a fit by
$\sigma(\phi) = \sigma_0 \left[1 + \beta \left\{1 - \cos\left(\gamma \left(\phi - \phi_0\right)\right)\right\}\right]$.
The minimum width of the histogram is obtained at the coordinate rotation angle of $0.482(3)^\circ$
and the lattice constant of 2.66(1) pixels on CCD plane
corresponding to the lattice constant of 266 nm.
From the same analysis of the other lattice axis,
we also obtain the coordinate rotation angle of  $-0.664(4)^\circ$ and lattice constant of 2.65(1) pixels on CCD plane
corresponding to the lattice constant of 266 nm.
Thus, the magnification ratio of our imaging system is 159.7(4)
and one pixel of our CCD camera corresponds to 100.2(3) nm on the objective plane.
These values are used for the deconvolution analysis of our images.
\Fref{fig:Image&PSF}(f) shows a raw image of the limited region of \Fref{fig:Image&PSF}(a)
with grid lines showing lattice separation and orientation.
In \fref{fig:Image&PSF}(g), we show a reconstructed atom distribution
where red circles and black dots represent the atoms and the lattice sites, respectively.
In \fref{fig:Image&PSF}(h), we also show the reconstructed atom distribution convoluted with the model PSF of equation \eref{eq:modelPSF},
which is compared with the raw image of \fref{fig:Image&PSF}(f).

An important aspect of QGM is the high-fidelity of the imaging process characterized by loss and hopping rates during the fluorescence imaging.
For this purpose, we take two successive images of the same atoms
with 400 ms exposure time and 300 ms delay between the two images,
and observe the change in the distribution.
We precisely tune the detuning $\delta_H$ of cooling molasses along the horizontal axes,
and evaluate the loss and hopping rates during the fluorescence image from the two successive images.
\Fref{fig:DistributionFraction}(a) and (b) show the results of reconstructed atom distributions at $\delta_H/2\pi = -394\ \mathrm{and}\ -206\ \mathrm{kHz}$, respectively.
For optimized parameters, we achieve loss rates of 6.5(1.8)\%
and hopping rates of 6.7(1.5)\%
for 400 ms exposures of clouds with fillings of $\sim$ 0.02 (see \fref{fig:DistributionFraction}(c)).
These rates, which include reconstruction errors, give a detection fidelity of 87(2)\%
for sparse clouds.

\section{Conclusion}
\label{sec:Conclusion}
In conclusion,
we demonstrate a bosonic $^{174}\mathrm{Yb}$ QGM for  a 2D optical lattice with a short lattice constant of 266 nm.
The atoms are preserved in the lattice sites during fluorescence imaging by narrow-line laser cooling which successfully combines Doppler cooling and sideband cooling.
The resulting temperature is
$T = 7.4(1.3)\ \mu\mathrm{K}$,
corresponding to a mean oscillation quantum number along the horizontal axes
0.22(4).
The PSF has a reasonably small width comparable to the ideal value, enabling the identification of the presence and absence of atoms by the deconvolution analysis. 
The high fidelity of the imaging process, which is an important aspect of QGM, is confirmed by the measurement of loss rate of 
6.5(1.8)\% 
and hopping rate of
6.7(1.5)\%
for 400 ms exposure time. 

While we perform the experiment with bosonic $^{174}\mathrm{Yb}$ atoms,
our method is applicable for a QGM for other  Yb isotopes, including fermionic $^{171}\mathrm{Yb}$ and $^{173}\mathrm{Yb}$.
In fact, our preliminary result shows that we can successfully obtain site-resolved images for fermionic $^{171}\mathrm{Yb}$ atoms.
In addition, the sideband cooling demonstrated in this work can be straightforwardly applied to other alkaline-earth atoms such as strontium,
especially for an optical lattice with magic wavelength.  
The realization of a QGM with enough fidelity for Yb atoms in a Hubbard-regime optical lattice opens up the possibilities
for studying various kinds of quantum many-body systems, and also long-range-interacting systems such as  Rydberg states.

\ack
We would like to thank 
K. Shibata for the setup of this experiment and 
T. Fukuhara for useful discussions.
This work was supported by the Grant-in-Aid for Scientific Research of JSPS
(Nos. 13J00122, 25220711, 26247064, 26610121)
and the Impulsing Paradigm Change through Disruptive Technologies (ImPACT) program.

\section*{References}

\providecommand{\newblock}{}

\end{document}